\documentclass[conference]{IEEEtran}
\IEEEoverridecommandlockouts
\usepackage{cite}
\usepackage{amsmath,amssymb,amsfonts}
\usepackage{algorithmic}
\usepackage{graphicx}
\usepackage{textcomp}
\usepackage{xcolor}
\usepackage{hyperref}
\def\BibTeX{{\rm B\kern-.05em{\sc i\kern-.025em b}\kern-.08em
    T\kern-.1667em\lower.7ex\hbox{E}\kern-.125emX}}

\makeatletter
\newcommand{\linebreakand}{%
  \end{@IEEEauthorhalign}
  \hfill\mbox{}\par
  \mbox{}\hfill\begin{@IEEEauthorhalign}
}
\makeatother
   
\begin{document}

\title{2SFGL: A Simple And Robust Protocol For Graph-Based Fraud Detection}

\author{\IEEEauthorblockN{Zhirui Pan}
\IEEEauthorblockA{\textit{Fudata Technology}\\
Shanghai, China \\
1970289663@qq.com}
\and
\IEEEauthorblockN{Guangzhong Wang}
\IEEEauthorblockA{\textit{Bank of Communications}\\
Shanghai, China \\
wang.guangzhong@bankcomm.com}
\and
\IEEEauthorblockN{Zhaoning Li}
\IEEEauthorblockA{\textit{Bank of Communications}\\
Shanghai, China \\
lizhaoning@bankcomm.com}
\linebreakand
\IEEEauthorblockN{Lifeng Chen}
\IEEEauthorblockA{\textit{Fudata Technology}\\
Shanghai, China \\
tianpu@fudata.cn}
\and
\IEEEauthorblockN{Yang Bian}
\IEEEauthorblockA{\textit{Fudata Technology}\\
Shanghai, China \\
douheng@fudata.cn}
\and
\IEEEauthorblockN{Zhongyuan Lai}
\IEEEauthorblockA{\textit{Fudan University}\\
Shanghai, China \\
abrikosoff@yahoo.com}
}

\maketitle

\begin{abstract}
Financial crime detection using graph learning improves financial safety and efficiency. However, criminals may commit financial crimes across different institutions to avoid detection, which increases the difficulty of detection for financial institutions which use local data for graph learning. As most financial institutions are subject to strict regulations in regards to data privacy protection, the training data is often isolated and conventional learning technology cannot handle the problem. Federated learning (FL) allows multiple institutions to train a model without revealing their datasets to each other, hence ensuring data privacy protection. In this paper, we proposes a novel two-stage approach to federated graph learning (2SFGL): The first stage of 2SFGL involves the virtual fusion of multiparty graphs, and the second involves model training and inference on the virtual graph. We evaluate our framework on a conventional fraud detection task based on the FraudAmazonDataset and FraudYelpDataset. Experimental results show that integrating and applying a GCN (Graph Convolutional Network) with our 2SFGL framework to the same task results in a 17.6\%-30.2\% increase in performance on several typical metrics compared to the case only using FedAvg, while integrating GraphSAGE with 2SFGL results in a 6\%-16.2\% increase in performance compared to the case only using FedAvg. We conclude that our proposed framework is a robust and simple protocol which can be simply integrated to pre-existing graph-based fraud detection methods.
\end{abstract}


\section{Introduction}
Identifying financial fraud and money laundering and generating the corresponding blacklist is a common method of risk management for financial institutions. Financial crime costs the industry annually up to tens of billions of dollars globally\cite{b1}, and its negative effects far exceed the monetary value itself. In order to avoid greater financial losses, financial institutions often spend a lot of human and financial resources to build a robust and stable IT system to prevent fraudulent activities. However, criminals could commit financial crimes across different financial institutions. Conventional fraud detection systems are usually hard-pressed to function well in the presence of cross-institutional data, as they need to integrate data originating from multiple sources and consisting of multiple modalities. From the point-of-view of financial institutions, integrating and utilizing user and transaction data distributed across different institutions to combat fraudulent activities has become an important challenge. 

Currently, these challenges are typically addressed by centralizing the data and employing traditional machine learning methods for identifying criminal behavior. However, the increased difficulty of achieving data centralization across institutions has led to a greatly elevated likelihood of false positives, against which an increasing amount of human resource needs to be utilized. The main reason for false-positive alerts is insufficient training data or insufficient variability in the training data structure\cite{b1}. In reality, graph structured data could be found everywhere (e.g., social networks, financial transaction networks, etc.). Graph learning uses various graph models to mine valuable information from graph structured data, often performing better than traditional machine learning methods, for tasks such as fraud and community detection\cite{b2}. More specifically, fraud detection methods that rely on anomaly detection requires the ability to isolate and identify clusters of nodes; these nodes could be of interest to banks due to their interconnections and could potentially signal the existence of a group of collaborating parties. Currently, most existing graph learning is designed for centralized scenarios, where graph data is stored centrally in the single graph database and model training is centralized. For example, most banks share common customers with other institutions. Hence a complete credit assessment could only be completed in collaboration with other institutions. One straightforward approach is to collect graph data from various institutions and merge them into a large graph, and then perform the relevant graph learning algorithms on the large graph. However, in the financial industry, graph data exist in form of islands\cite{b3}, and hence it is unrealistic to collect these graph data in a centralized manner because of privacy concerns, the independence of each organization’s graph data, and competition with each other. Therefore, to solve the problem of false-positive alerts, it is crucial to organize graph data from different institutions for combined calculations, while simultaneously ensuring data privacy and security. 

Two main challenges should be solved in research on federated graph learning:
\begin{itemize}
    \item Heterogeneity: In the process of federated graph learning, the graph data of each client is likely to be highly Non-Independent Identically Distributed (Non-IID), i.e. the probability distribution of each data point is unlikely to be the same and each point unlikely to be independent from one another\cite{b4}; on the contrary, highly correlated data is to be expected in large-scale structured data such as those commonly found in graph data. In which case the final trained global model is likely to be unsatisfactory due to the large differences between the locally trained models of the clients;
    \item Complementarity: There are a large number of duplicate vertices in the graph data between clients, and the graph structure (vertex, vertex-to-vertex connections) of each client is not comprehensive due to the inability to share and aggregate the original graph data of each party.
\end{itemize}

In this paper, we propose a novel two-stage approach to federated graph learning (2SFGL) for further solving the above-proposed problem. In the first stage of 2SFGL, each client normalizes their respective edge values with respect to local graph data and then virtually fuses them into each other, thus alleviating the heterogeneity problem and leveraging complementarity. The second stage of 2SFGL is to apply relevant graph learning algorithms, such as the label propagation algorithm (LPA), PageRank or Graph Neural Network(GNN) on the fused graph. The 2SFGL procedure proposed avoids the training of local models and hence greatly reduces the problem of unsatisfactory learning results due to heterogeneity and complementarity (e.g., instability of the trained global model).

\section{Background}

\subsection{Privacy-preserving Strategies}

Privacy is one of the most important goals of federated learning. Data private collaborative learning introduces additional restrictions to the training process over that of data-shared learning as the computational process is not identical. In this section, we briefly review and compare different privacy techniques, which can be combined with federated learning. The methods of differential privacy\cite{b5} involve adding noise to the data or using generalization methods to obscure certain sensitive attributes until the third party cannot distinguish the individual. Secure Multi-party Computation (SMPC) naturally involves multiple parties, with each party knowing nothing except its input and output. It is possible to build a security model with SMPC under lower security requirements in exchange for efficiency\cite{b6}. Homomorphic Encryption has also been adapted to protect user data privacy through parameter exchange under the encryption mechanism during machine learning. Recent works are widely used and polynomial approximations need to be made to evaluate non-linear functions in machine learn algorithms\cite{b7}\cite{b8}.

\subsection{Federated Learning}
\begin{figure}[htbp]
\centerline{\includegraphics[width=0.4\textwidth]{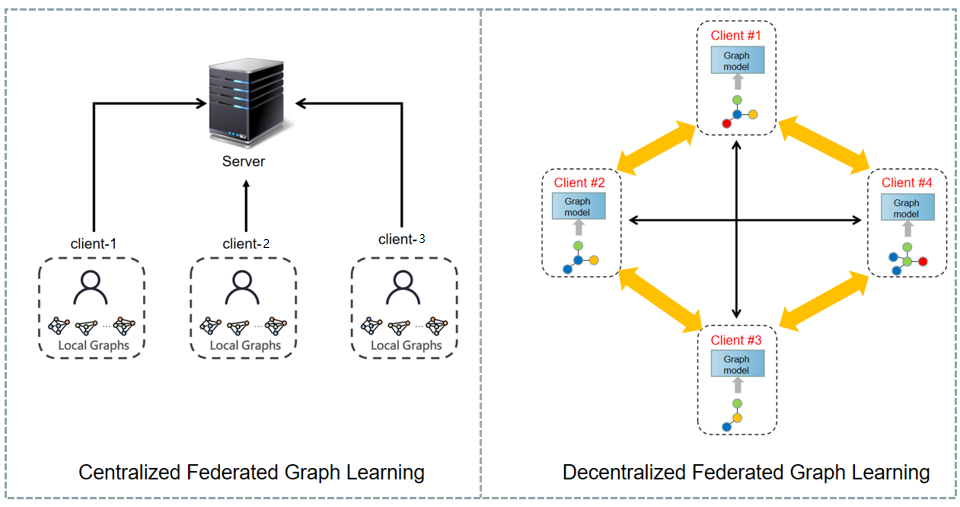}}
\caption{The left figure represents centralized federated graph learning, and the right figure represents decentralized federated graph learning.}
\label{fig1}
\end{figure}

Federated learning\cite{b9} is a novel technique used to train models for machine learning based on datasets distributed across different institutions, meanwhile ensuring that data privacy is not compromised. Federated learning is classified into horizontal and vertical federated learning, as well as federated transfer learning. Horizontal FL (HFL) is the case of federated learning where all users across the network trains their respective models with data having the same set of features, while vertical FL (VFL) is the setting where each user across the network trains a common model using different datasets with varying features\cite{b10}. Up to now, many improvement efforts have been devoted to addressing various problems of federated learning, including reducing network bandwidth traffic\cite{b11} and more effective protection of data privacy\cite{b12}. Federated learning is applicable in finance, healthcare, and other fields. Recent research on federated learning on graph data can be understood as either centralized or decentralized federated graph learning. Centralized graph federated learning implies the existence of multiple clients and a central server, which combines multiple clients for model training and shares the global model with the clients, while in decentralized federated graph learning, multiple participants collaborate to train the same model without a centralized server to control the entire model update. In Fig.~\ref{fig1}, the left side of the figure represents the centralized federated graph learning process with $K$ clients and one central server. Each client trains a graph model based on local graph data, and the server receives the local model parameters or gradients sent by all clients to update the global model and then distributes the global model to each client for the next batch of training. However, deciding who should be the central server is a difficult process. The right side of the figure represents the decentralized federated graph learning process. Lalitha et al.\cite{b13} formally describes the fully decentralized federated learning problem and proposed an efficient distributed training method. Zhao et al.\cite{b14} removed the centralized server and updated the parameters for federated learning between clients. FedAvg\cite{b9} uses a federated learning approach that aggregates the average of the respective local models. The ultimate goal of federated learning is to obtain a more generalized global framework that is suitable for all clients, while at the same time preserving clients' privacy. However, in some recent studies\cite{b15,b16} when the local data of each client is Non-IID, existing federated learning methods are difficult to obtain a better global generalization model by training.

\subsection{Graph-based Machine Learning}
Graph learning is a machine learning method that uses graph data for classification, clustering, and feature extraction. A range of graph learning is used to determine graph features, such as connectivity, centrality, and community discovery. Graph features can be combined with non-graph features and training data labels are provided to define as supervised machine learning. If training data is not provided with labels, then the problem is an unsupervised machine learning problem, we can apply machine learning algorithms such as clustering, dimensionality reduction, or PageRank. Recently, there have been many advances in scalable graph computation for billion-scale or even trillion-scale graphs\cite{b17}\cite{b18}, so it is reasonable to expect that this approach would remain practical, even for large graphs. Akoglu et al.\cite{b19} studied graph-based anomaly detection. Moloy et al.\cite{b20} used PageRank algorithm for fraud detection. Liu et al.\cite{b21} used graph embedding methods to detect financial criminal activities, such as anti-money laundering (AML). Recently, some works\cite{b22}\cite{b23} have studied GNN-based fraud detection. Weber et al.\cite{b24} used graph convolutional networks (GCN) to detect money laundering. Pareja et al.\cite{b25} explored dynamic considerations in graph networks, which is essential in fraud detection applications. Dou et al.\cite{b26} proposed solutions to expose camouflaged fraudsters via GNNs. We note that these previous works focus more on local graph features, while our current work concentrates on global graph features across multiple institutions.

\section{Related works}
In this part, we survey related work on SMPC on graphs, federated learning on GNNs, and graph-based fraud detection.

\subsection{Secure Multi-party Computations on Graphs}
The importance of secure multi-party computations has increased in the past decade\cite{b27}. It allows several parties to build a model on their pooled data to increase utility while not explicitly sharing data with each other. Secure two-party computation was first investigated by Yao and was later generalized to multiparty computation\cite{b28}. Secure multi party computation based privacy preserving techniques are usually adopted to privacy through secret sharing and homomorphic encryption scheme for the original graph data. On the other hand, secure multi-party computations on graphs, such as breath-first-search, leader election, secure sum on graphs, secure minimum and maximum search of all inputs on graphs, and secure vertex coloring, as all much more rarely studied. Currently, there is little research on SMPC-based graph computation. We applied SMPC to the virtual fusion process of the graph.

\subsection{Federated Learning on Graph Neural Networks}
Recently, some works have trained graph-data models in federated learning settings, mainly by combining federated learning with graph neural networks, and several federated graph neural networks have been proposed. For example, Zheng et al.\cite{b29} designed a new federated learning framework for graph convolutional neural networks, Wu et al.\cite{b30} designed a federated GNN framework for privacy-preserving recommendations, Jiang et al.\cite{b31} proposed a dynamic representation method for learning objects from multi-user graph sequences. We train the model on the respective party-based virtual fusion graphs by FL and hence leveraging heterogeneity and complementarity.

\subsection{Graph-based Fraud Detection}
Financial fraud detection is based on behavioral data from financial platforms to detect malicious accounts, defaulting users, and fraudulent transactions. GEM et al.\cite{b32} adaptively learns discriminative embeddings from heterogeneous account-device graphs for malicious account detection. Semi-GNN\cite{b33} is a semi-supervised GNN model with a hierarchical attention mechanism for explainable fraud prediction. FdGars\cite{b34} is a graph convolutional network approach for fraudster detection in an online app review system. GraphConsis\cite{b34} investigates the context, feature, and relation inconsistency problem in graph-based fraud detection. CARE\cite{b35} enhances the GNN aggregation process against camouflage for opinion fraud detection. These works are based on the FL-on-GNN approach to fraud detection; we also employ a similar approach in this paper. However, in our case we perform a graph virtual fusion algorithm before graph federal learning for model training, thus promoting fast convergence of models.

\section{Two-Stage Approach to Federated Graph Learning for Fraud Detection}

\begin{figure}[htbp]
\centerline{\includegraphics[width=0.45\textwidth]{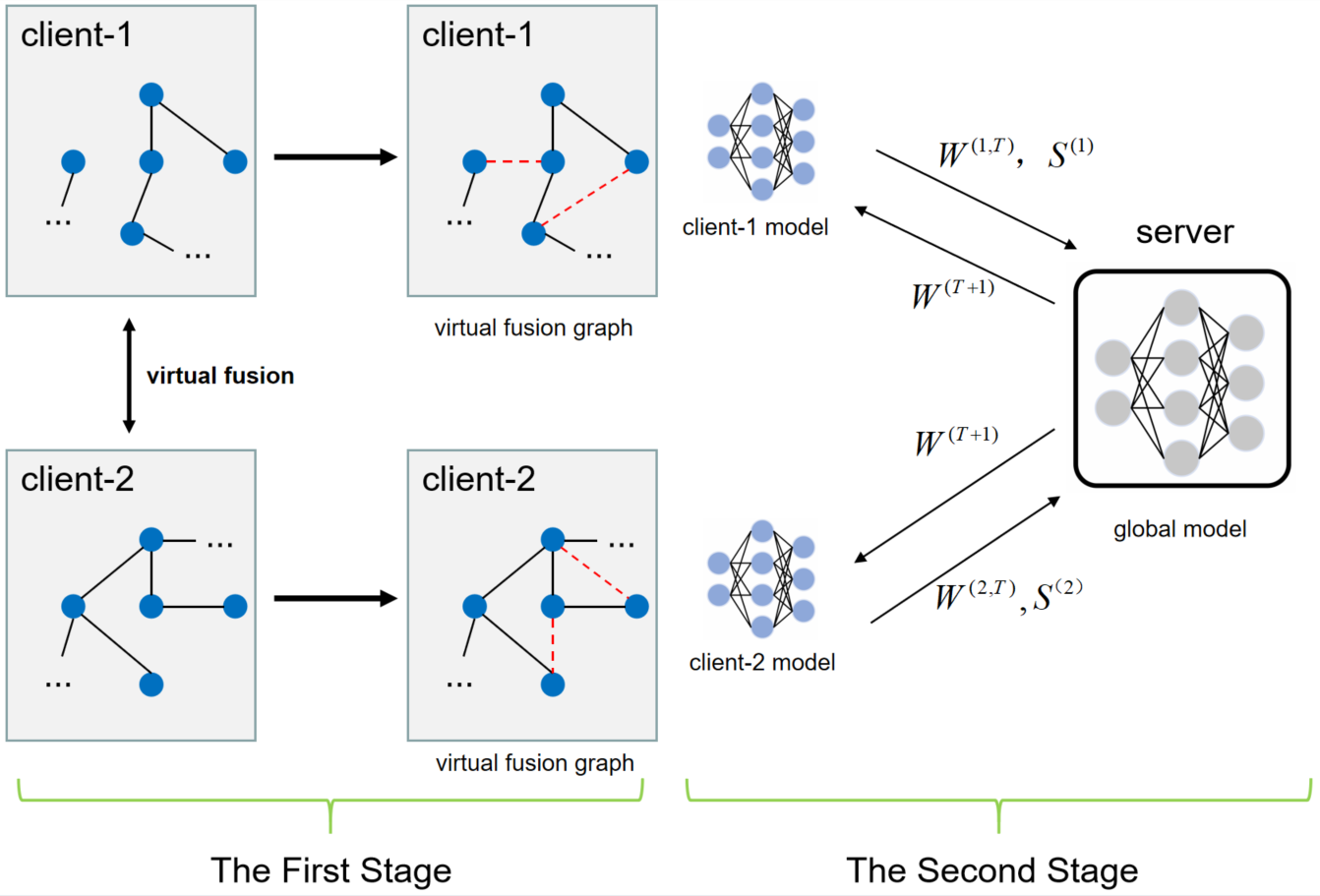}}
\caption{Overview of 2SFGL. The first stage of the 2SFGL is virtual fusion by the respective client, and the second stage is FedAvg on the virtual fusion graph of the respective client}
\label{figover}
\end{figure}

In this part, we present a new federated learning framework that combines graph data across different clients for a fast virtual fusion approach to enrich graph data from different clients. And then using the FedAvg method on the virtual fusion graphs to train GNN model. An overview of the framework is shown in Fig.~\ref{figover}.

\subsection{The First Stage of 2SFGL}

In the first step of 2SFGL, we apply Private Set Intersection (PSI)\cite{b36} to ensure the privacy and security of graph data. PSI is a cryptographic technique in SMPC that allows each party involved in the computation to compute the intersection of their data without obtaining additional information from the other party. The PSI technique allows multiple participants to compute the common vertices of graph data without revealing local source information. Specifically, PSI acquires two or even more parties of customers with common vertices. Each client separately finds the edge relations of vertices common to their respective graphs. We cannot directly send edge information between common vertices to other participants, so we first normalize edge values as follows:

\begin{align}
{\rm{N}}_{ij}^{} = \left\{ {\begin{array}{*{20}{c}}
{\begin{array}{*{20}{c}}
{\frac{{{E_{ij}}}}{{\sum\limits_{k = 1}^n {{E_k}} }}}&{}&{{E_{ij}} > 0}
\end{array}}\\
{\begin{array}{*{20}{c}}
0&{}&{}&{{E_{ij}} = 0}
\end{array}}
\end{array}} \right.
\end{align}

where ${E_{ij}}$ denotes the value of edges between vertices ${V_i}$ and ${V_j}$, $n$ is the total number of edges connected to vertices ${V_i}$, $\sum\limits_{k = 1}^n {{E_{k}}}$ denotes the sum of the values of the edges connected to vertices ${V_i}$, and ${{\rm{N}}_{ij}}$ denotes the normalized result of the value of the edge relationship between vertices ${V_i}$ and ${V_j}$. Then differential privacy is applied to ${{\rm{N}}_{ij}}$, to avoid the exposure of sensitive individual information.

\begin{figure}[htbp]
\centerline{\includegraphics[width=0.45\textwidth]{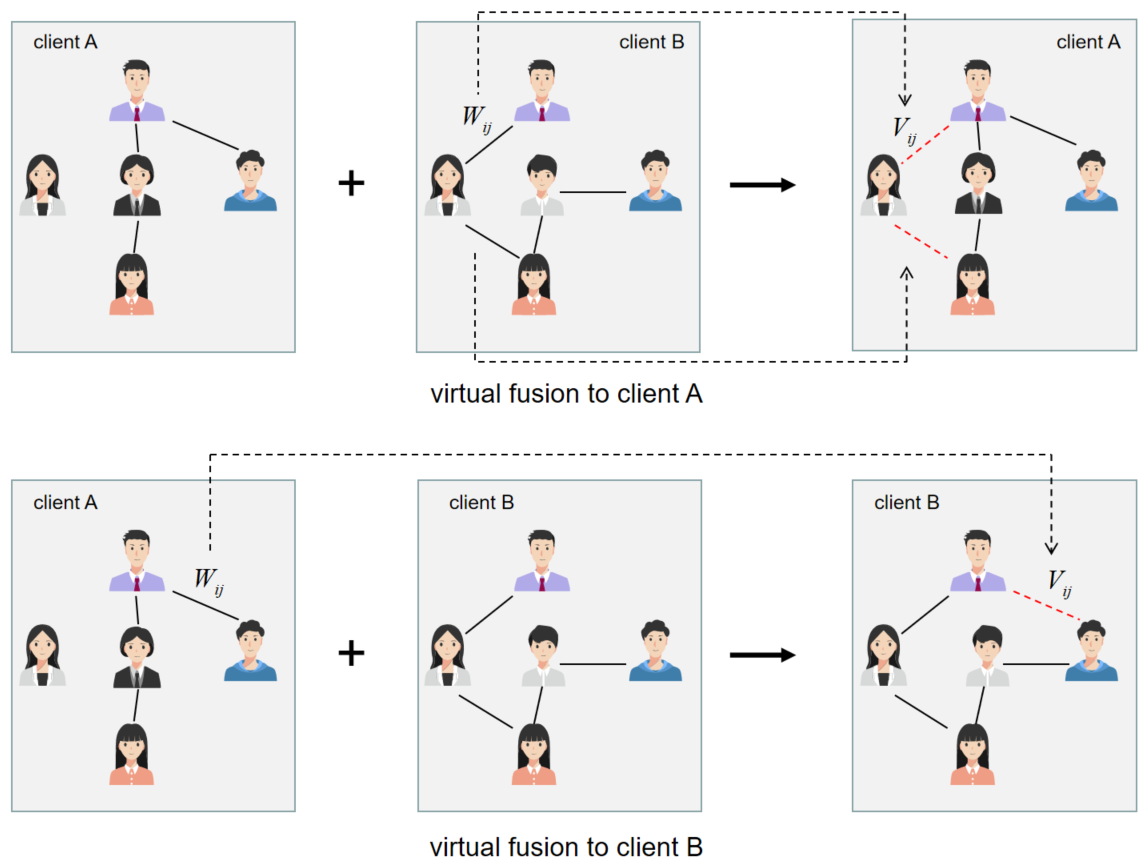}}
\caption{One hop virtual fusion.}
\label{fig2}
\end{figure}

\begin{figure}[htbp]
\centerline{\includegraphics[width=0.45\textwidth]{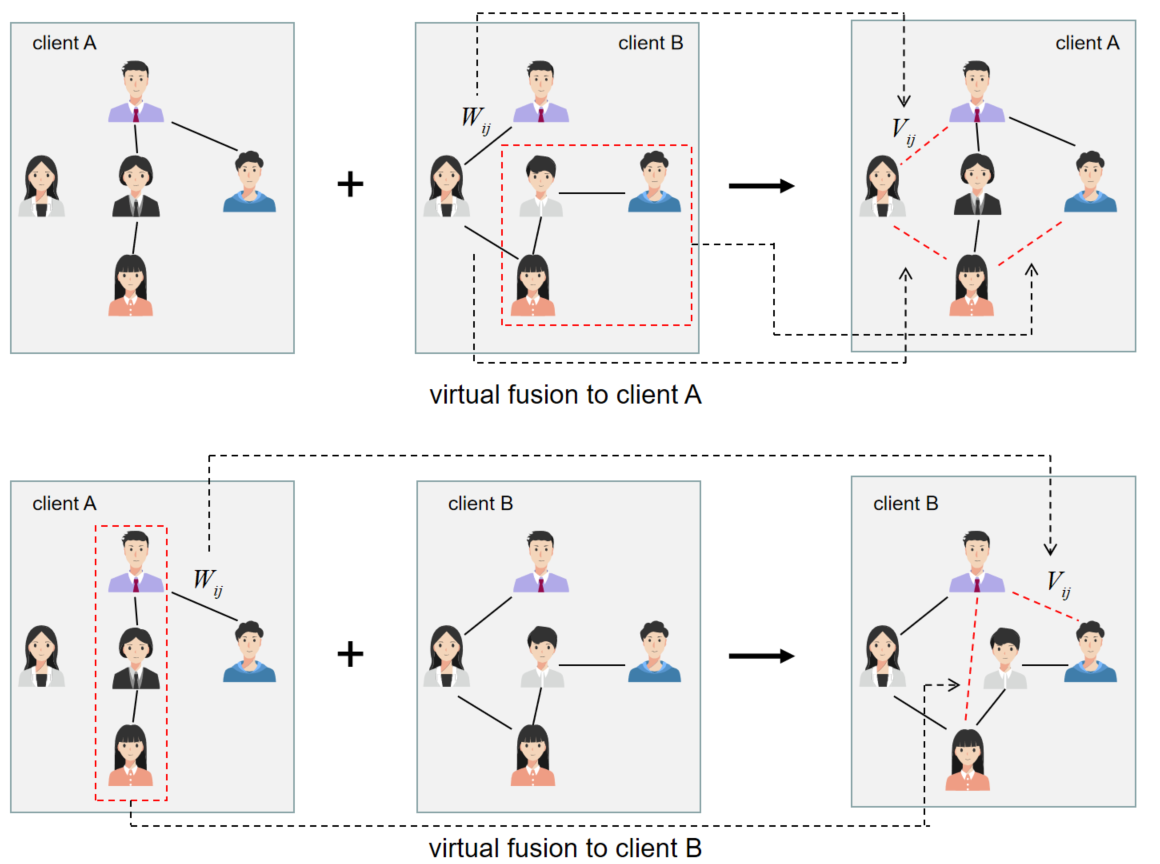}}
\caption{Two hop virtual fusion.}
\label{fig3}
\end{figure}

Each specific client sends the normalized results of all common vertices to the other clients, and the updated common vertex edge value is calculated using information from PSI:
\begin{align}
{{\rm{E}}^{new}_{ij}} = \left\{ {\begin{array}{*{20}{c}}
{\frac{{{{\rm{N}}_{ij}}}}{{1 - {{\rm{N}}_{ij}}}}\sum\limits_{k = 1}^n {{E_k}} \begin{array}{*{20}{c}}
{}&{}&{{{\rm{N}}_{ij}} < \lambda }
\end{array}}\\
{\frac{\lambda }{{1 - \lambda }}\sum\limits_{k = 1}^n {{E_k}} \begin{array}{*{20}{c}}
{}&{}&{}&{{{\rm{N}}_{ij}} \ge \lambda }
\end{array}}
\end{array}} \right.
\end{align}
where ${\rm{E}}^{new}_{ij}$ is the updated value of the edge relationship between vertices ${V_i}$ and ${V_j}$, ${{\rm{N}}_{ij}}$ denotes the normalized result of the value of the edge relationship between vertices ${V_i}$ and ${V_j}$ sent from other clients. $\sum\limits_{k = 1}^n {{E_k}}$ denotes the sum of the values of the vertex ${V_i}$ neighboring edges, and $\lambda$ is the threshold to be set by user of the federated graph learning framework. The threshold value indicates the importance of the edge relationship of the client’s graph data. In this paper, the threshold value of 0.5 is used, which indicates that the client-side edge relationship values of two graphs from separate parties are at least of equal importance. We call the augmented virtual graph obtained after performing the computations detailed above a~\emph{virtual fusion} graph, and the process is a decentralized one. Under our framework, we can perform multiple~\emph{hop} virtual fusion. Multiple hop virtual fusion can be chosen from one, two, or three hop virtual fusion, where one hop virtual fusion is a virtual fusion of edges with one hop, and two hop virtual fusion is a virtual fusion of edges with one hop and two hops respectively. Fig.~\ref{fig2} shows a one hop virtual fusing process, and Fig.~\ref{fig3} shows a two hop virtual fusing process.

\subsection{The Second Stage of 2SFGL}
After the first stage of 2SFGL, the relevant algorithms are performed on the virtual fusion graph. In addition, the stage is a centralized process. In this paper, we use conventional graph neural network algorithms such as GCN and GraphSAGE, to verify the performance of our federated graph learning framework on the FraudAmazonDataset. In brief, the GCN is defined as:
\begin{align}
\mathrm{H}=\tilde{D}^{-\frac{1}{2}} \tilde{A} \tilde{D}^{-\frac{1}{2}} X W
\end{align}
where ${\rm{X}} \in {R^{N \times F}}$ is the input matrix, ${\rm{H}} \in {R^{N \times {F^{'}}}}$ is the convolved matrix, and ${\rm{W}} \in {R^{F \times {F^{'}}}}$ is the parameter. ${F}$ and ${F^{'}}$ are the dimensions of the input and the output, respectively. GraphSAGE is a general inductive framework which generates embeddings by sampling and aggregating features from a node’s local neighborhood\cite{b37}:
\begin{align}
{\begin{array}{l}
{\rm{h}}_{{{\rm N}_v}}^{{\rm{t + 1}}} = {\rm{AG}}{{\rm{G}}_{t + 1}}(\{ {\rm{h}}_u^t,\forall u \in {{\rm N}_v}\} ),\\
{\rm{h}}_v^{t + 1} = \sigma ({{\rm{W}}^{t + 1}} \bullet [{\rm{h}}_v^t||{\rm{h}}_{{{\rm N}_v}}^{{\rm{t + 1}}}]).
\end{array}}
\end{align}
where ${\rm{AG}}{{\rm{G}}_{t + 1}}$ is an aggregation function\cite{b37}, and ${\sigma}$ is an activation function, such as the sigmoid function.

In the second stage, we follow the FedAvg algorithm. FedAvg has become the de facto FL algorithm where clients communicate with the central server at each epoch. The second stage of the 2SFGL is shown in Fig.~\ref{figover}. In the ${T}$-th epoch, a client sends local model parameters ${W^{(k,T)}}$ to the central server, where k is the kth client, and ${T}$ is the ${T}$-th epoch. The central server averages all the updates from the clients to obtain the global update ${W^{(T{\rm{ + }}1)}}$ which is broadcast to all clients in the (${T}$+1)-th epoch. In addition, the full second phase is a centralized process.
\begin{align}
{\begin{array}{l}
{W^{(T{\rm{ + }}1)}}{\rm{ = }}\sum\nolimits_{k = 1}^N {\frac{{{S^{(k)}}}}{{\sum\nolimits_{i = 1}^N {{S^{(i)}}} }}} {W^{(k,T)}}
\end{array}}
\end{align}

\section{Experiments}
\subsection{Dateset and Settings}
We use FraudAmazonDataset\cite{b38} and FraudYelpDataset\cite{b39} to establish the efficacy of 2SFGL on the fraud detection task. The FraudAmazonDataset includes product reviews under the Musical Instrument category. The vertices in the constructed graph for the FraudAmazonDataset are users with 100-dimension features and contains three types of connections: 1) $U-P-U$, connecting users reviewing at least one common product; 2) $U-S-U$, connecting users having at least one same star rating within one week, and 3) $U-V-U$, connecting users with top 5\% common review texts (as measured by TF-IDF) among all users. The FraudAmazonDataset can be used for the common fraud detection task, which is to find spam comments on online platforms. The FraudYelpDataset includes hotel and restaurant reviews filtered (spam) and recommended (legitimate) by Yelp. A spam review detection task can be conducted, which is a binary classification task. The nodes in the graph of YelpChi dataset are reviews with 100-dimension features and have three relations: 1) $R-U-R$ denotes the reviews posted by the same user; 2) $R-S-R$ denotes the reviews under the same product with the same star rating; 3) $R-T-R$ denotes the reviews under the same product posted in the same month.

In dataset sampling, we ensure that the ratio of positive to negative samples is in the range of 1:2 to 2:1, while ensuring that the train, test ratio are set to be 60\%, 40\% respectively. The parameters of GCN and GraphSAGE are optimized with Adam optimizer, the learning rate is set to be 0.005, and add only one hidden layer, the number of neurons in its hidden layer is 64. Specifically, the neighborhood size is set to 5 in GraphSAGE.

\subsection{Baseline Performance Analysis}
In the first stage of 2SFGL, we apply virtual fusion on FraudAmazonDataset’s three graph connections ($U-P-U$, $U-S-U$, $U-V-U$) within one hop relationship. Likewise we virtual fusion FraudYelpDataset's thre graph connections ($R-U-R$, $R-S-R$, $R-T-R$). In the second stage of 2SFGL, We use FedAvg method for model training of GCN and GraghSage on virtual fusion of multi-party graphs. For a fair comparison, we compare 2SFGL with a Federated Learning approach using only FedAvg. Also to demonstrate the effect of 2SFGL, GCN and GraphSAGE were also used separately on the three relational single local graph. We measure the performance using generic metrics, namely Accuracy, Macro-F1, AUC, and GMean\cite{b40}\cite{b41}.

\begin{figure}[htbp]
\centerline{\includegraphics[width=0.5\textwidth]{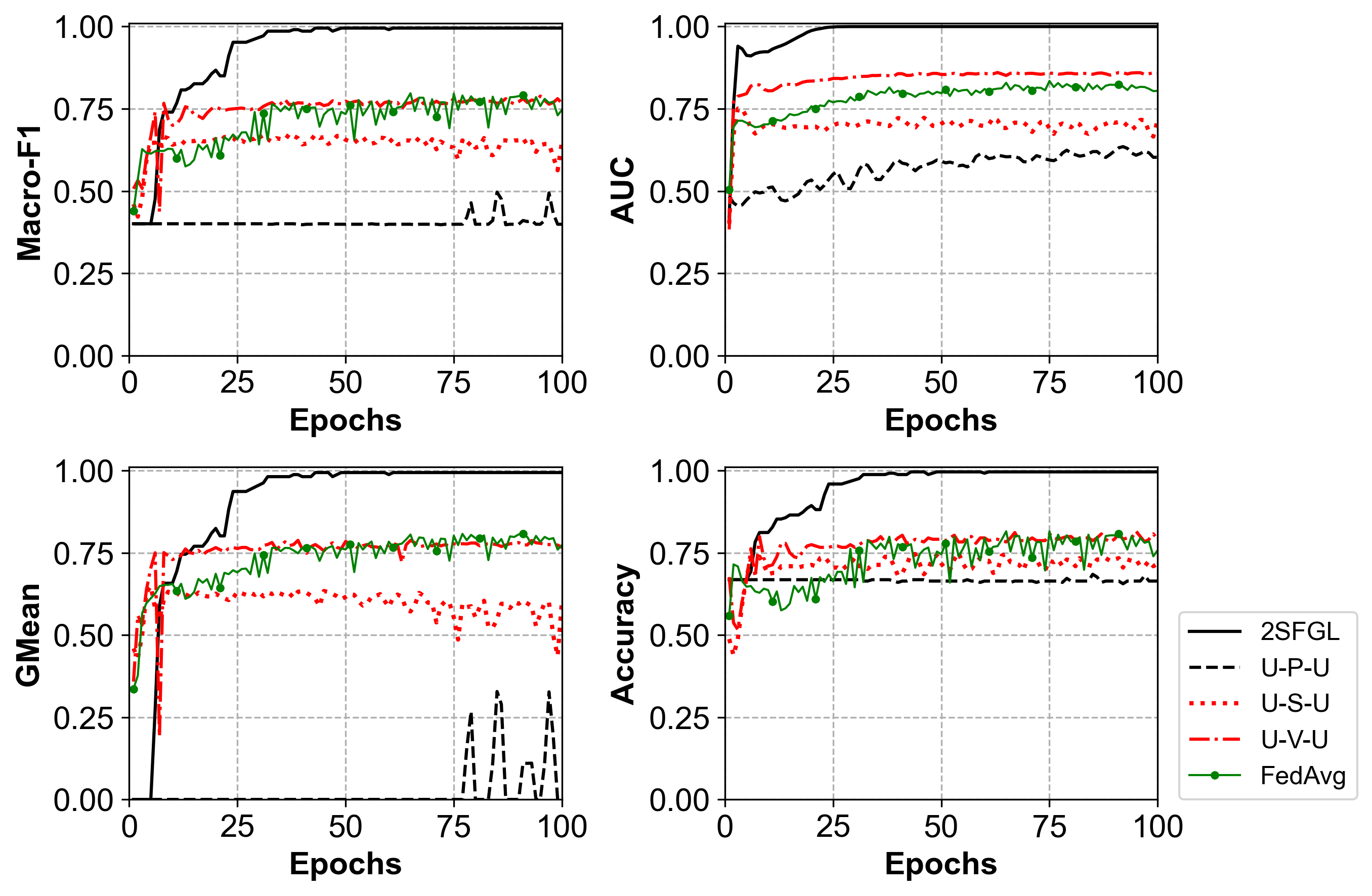}}
\caption{Sensitivity analysis result of GCN with respect to epochs using FraudAmazonDataset.}
\label{fig5}
\end{figure}

\begin{figure}[htbp]
\centerline{\includegraphics[width=0.5\textwidth]{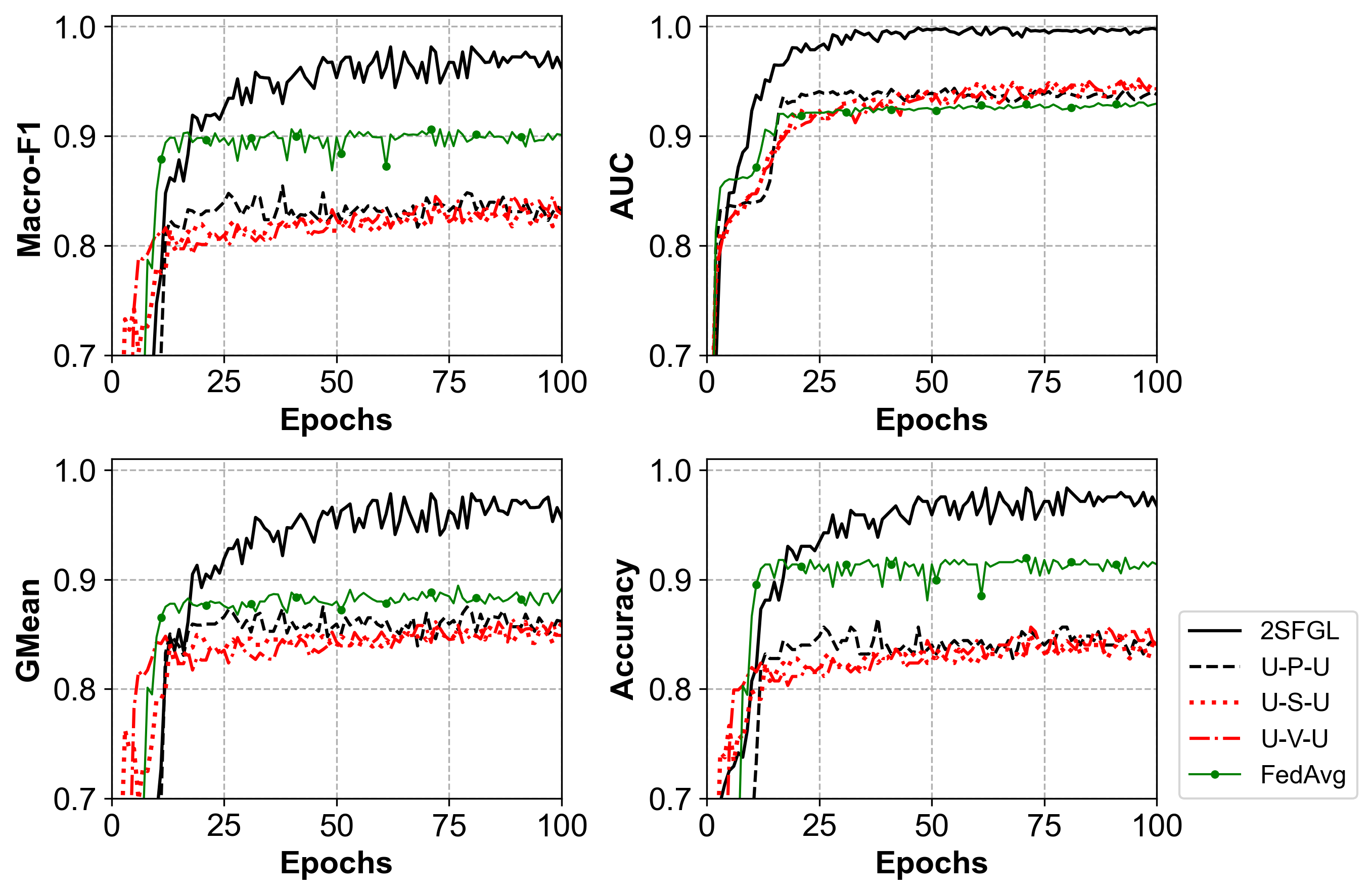}}
\caption{Sensitivity analysis result of GraphSAGE with respect to epoches using FraudAmazonDataset.}
\label{fig6}
\end{figure}

Fig.~\ref{fig5} shows the performance of using the FraudAmazonDataset on virtual fusion graphs, FedAvg only, and three relational single local graphs using GCN, respectively. From Fig.~\ref{fig5}, the solid black line represents results using the virtual fused graph on the 2SFGL. The green dotted solid line represents results using FedAvg. The black dashed line represents result using the dataset of U-P-U relationship. The red dotted line represents metrics result using U-S-U relationship. The red dash dot represents metrics result using U-V-U relationship. Fig.~\ref{fig6} shows the performance of using GraphSAGE, differing from Fig.~\ref{fig5} in that the metrics are performed using GCN. We can observe that the model obtained from the GCN and GraphSAGE of metrics using 2SFGL are better than using FedAvg and performing GNN inference on the single local graphs. We provide a brief analysis of the different metrics, Fig.~\ref{fig5} shows that when the model is trained using only U-P-U graph data, the prediction results predict almost all negative samples as positive samples, and the True Positive (TP) hardly increases, which leads to almost zero GMean and almost a constant Macro-F1 metric for the GCN algorithm. The sharp increase in AUC when training using only U-P-U data is shown in Figure 6 because the model training has not yet converged. In addition, GraphSAGE is trained using random sampling and aggregation methods, while GCN utilizes the adjacency matrix of the graph, also because this difference leads to different results, which are also reflected in some other studies\cite{b42}.

\begin{figure}[htbp]
\centerline{\includegraphics[width=0.5\textwidth]{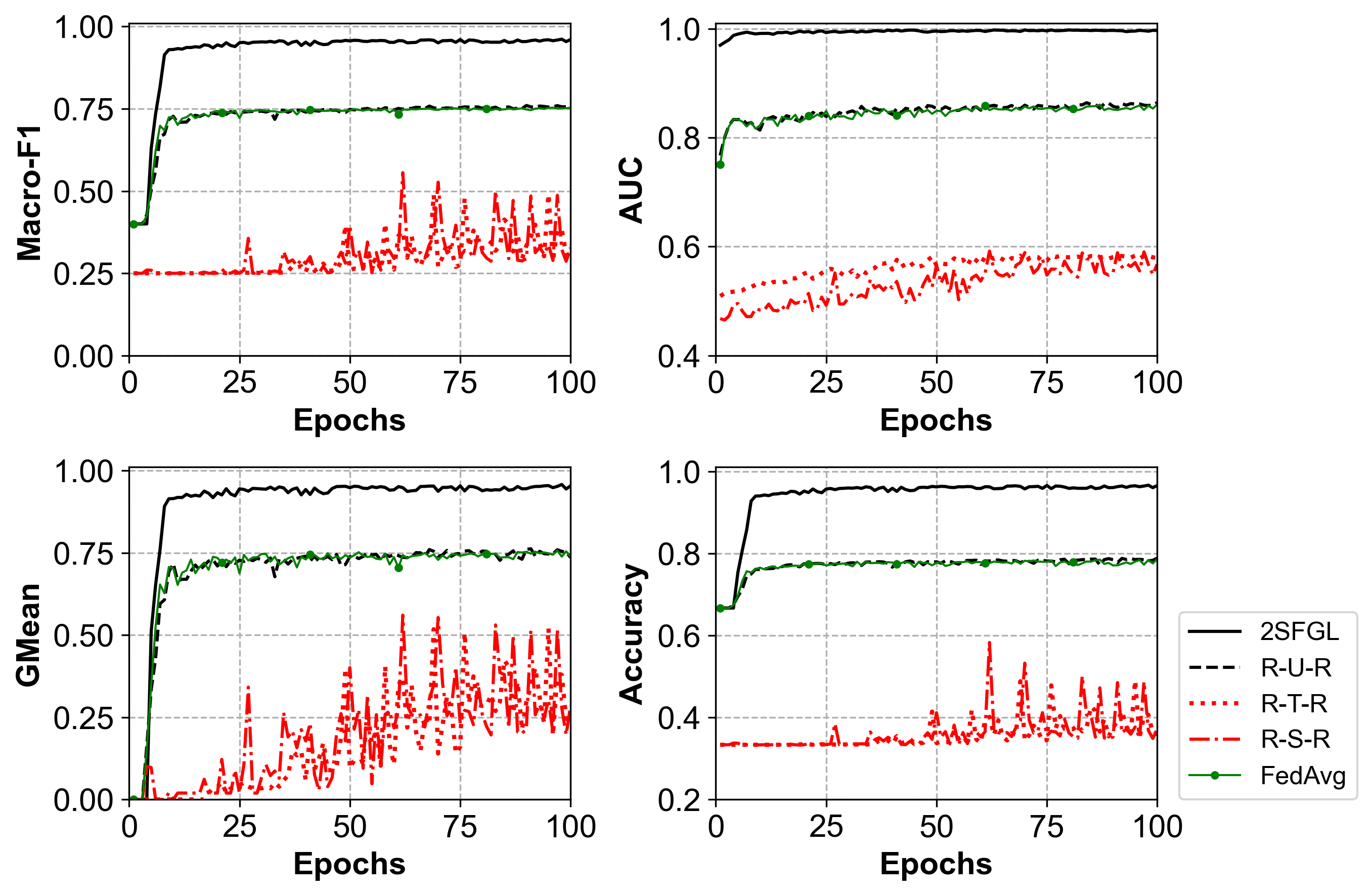}}
\caption{Sensitivity analysis result of GCN with respect to epoches using FraudYelpDataset.}
\label{fig7}
\end{figure}

\begin{figure}[htbp]
\centerline{\includegraphics[width=0.5\textwidth]{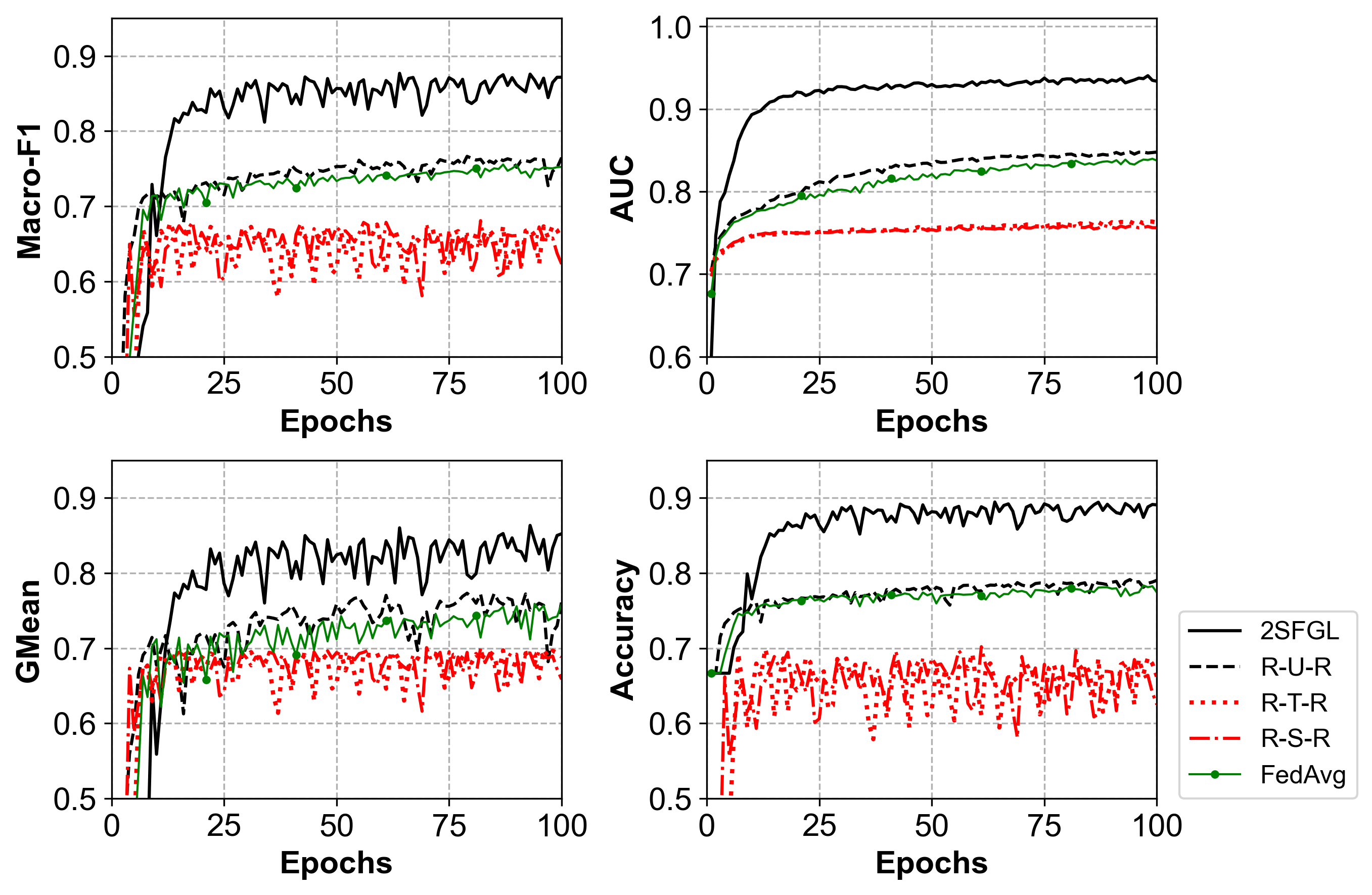}}
\caption{Sensitivity analysis result of GraphSAGE with respect to epoches using FraudYelpDataset.}
\label{fig8}
\end{figure}

Similarly, using the FraudYelpDataset for comparison. Fig.~\ref{fig7} shows the performance of using the FraudYelpDataset on virtual fusion graphs, FedAvg only, and three relational single local graphs using GCN, respectively. Differing from Fig.~\ref{fig8} in that the metrics are performed using GCN. Overall, the metrics results using 2SFGL are higher than only using the FedAvg method.

In Table I, we provide the statistical averages of the different metrics for Fig.~\ref{fig5} and Fig.~\ref{fig6} from the 60th epochs to the 100th epochs of training using FraudAmazonDataset. The results from the metrics of 2SFGL with using virtual fusion in the First stage and using GCN-based FedAvg in the second stage is 6\%-30.2\% higher than that of using only GCN-based FedAvg. In Table II, The only difference from Table I is that the dataset used is the FraudYelpDataset.
Similar results, using 2SFGL is 12\%-28.4\% higher than using only GCN-based FedAvg.

\begin{table}[htbp]
\caption{The statistical average of the metrics using FraudAmazonDataset}
\begin{center}
\begin{tabular}{|c|c|c|c|c|c|}
        \hline
        GNN & Relationships & Macro-F1 & AUC & GMean & Accuray  \\ \hline
        ~ & 2SFGL & 0.99 & 1.0 & 0.99 & 0.99  \\ 
        ~ & FedAvg & 0.76 & 0.81 & 0.78 & 0.77  \\ 
        GCN & U-P-U & 0.41 & 0.60 & 0.04 & 0.66  \\ 
        ~ & U-S-U & 0.64 & 0.70 & 0.59 & 0.72  \\ 
        ~ & U-V-U & 0.77 & 0.86 & 0.77 & 0.79  \\ \hline
        ~ & 2SFGL & 0.97 & 0.99 & 0.96 & 0.97  \\ 
        ~ & FedAvg & 0.90 & 0.93 & 0.88 & 0.91  \\ 
        GraphSAGE & U-P-U & 0.83 & 0.93 & 0.86 & 0.84  \\ 
        ~ & U-S-U & 0.82 & 0.94 & 0.85 & 0.83  \\ 
        ~ & U-V-U & 0.83 & 0.94 & 0.85 & 0.84  \\ \hline
\end{tabular}
\label{tab1}
\end{center}
\end{table}

\begin{table}[htbp]
\caption{The statistical average of the metrics using FraudYelpDataset}
\begin{center}
\begin{tabular}{|c|c|c|c|c|c|}
        \hline
        GNN & Relationships & Macro-F1 & AUC & GMean & Accuray  \\ \hline
        ~ & 2SFGL &          0.96 & 1.0 & 0.95 & 0.96  \\ 
        ~ & FedAvg &         0.75 & 0.85 & 0.74 & 0.78  \\ 
        GCN & R-U-R &        0.75 & 0.86 & 0.74 & 0.78  \\ 
        ~ & R-T-R &          0.32 & 0.58 & 0.26 & 0.37  \\ 
        ~ & R-S-R &          0.33 & 0.55 & 0.27 & 0.38  \\ \hline
        ~ & 2SFGL &          0.86 & 0.93 & 0.83 & 0.88  \\ 
        ~ & FedAvg &         0.74 & 0.83 & 0.73 & 0.77  \\ 
        GraphSAGE & R-U-R &  0.75 & 0.84 & 0.75 & 0.78  \\ 
        ~ & R-T-R &          0.65 & 0.76 & 0.68 & 0.66  \\ 
        ~ & R-S-R &          0.65 & 0.76 & 0.68 & 0.66  \\ \hline
\end{tabular}
\label{tab2}
\end{center}
\end{table}

\section{Conclusion}
In this paper, we propose 2SFGL, a novel two-stage approach (the first stage is the virtual fusion of multiparty graphs, the second is model training and inference in the virtual graph using FedAvg) for increasing the accuracy of financial crime identification. The 2SFGL is implemented by combining graph-based learning and federated learning. Using FraudAmazonDataset to GCN on 2SFGL, metrics are 23.5\%-30.2\% better than GCN on only using FedAvg. While to GraphSAGE, improve the metrics by 6\%-9.1\%. Similar, Using FraudYelpDataset to GCN on 2SFGL, metrics are 17.6\%-28.4\% better than GCN on only using FedAvg. While to GraphSAGE, improve the metrics by 12\%-16.2\%.

we are actively working on pilots, especially in banking institutions in the financial system. We will organize multiple financial institutions (e.g., banks) and communications department (e.g., operators) to participate in leveraging the federated graph learning framework for accurate detection of financial crimes.

\section*{Acknowledgment}

\end{document}